\documentclass[11pt]{article}
\usepackage{times}
\usepackage{sidecap}
\usepackage{geometry}
\usepackage{floatrow}
\usepackage{float}
\usepackage{hyperref}
\usepackage{graphics,graphicx}
\usepackage{subcaption}
\usepackage{epsf,wrapfig}
\usepackage{ulem}
\usepackage{natbib,bm}
\usepackage{paralist}
\usepackage{txfonts}
\usepackage{booktabs}
\usepackage{siunitx}
\usepackage{comment}
\let\olditem\item
\renewenvironment{thebibliography}[1]{%
  \section*{\refname}
  \let\par\relax
  \renewcommand{\item}[1][]{\olditem[\textbullet]}%
  \inparaenum}{\endinparaenum}

\newcounter{aff}

\DeclareGraphicsExtensions{.jpg,.pdf,.png,.eps}

\usepackage{caption}
\usepackage[dvipsnames]{xcolor}

\usepackage{tcolorbox}

\bibpunct{(}{)}{;}{a}{}{,}

\usepackage{soul}

\definecolor{darkcyan}{rgb}{0.0, 0.0, 0.7}

\definecolor{boxcolor}{RGB}{193, 229, 255}
\definecolor{figboxcolor}{RGB}{193, 229, 255}
\definecolor{boxcolorgray}{RGB}{227, 227, 227}
\definecolor{darkcyan}{rgb}{0.0, 0.0, 0.7}

\sethlcolor{boxcolorgray}
\colorlet{boxcolorapricot}{Apricot}
\colorlet{boxcoloremerald}{Emerald}

\definecolor{Teal}{HTML}{00A6A6}
\definecolor{Cyan}{HTML}{00A9CE}
\definecolor{Jade}{HTML}{00A878}
\definecolor{Azure}{HTML}{007ACC}
\definecolor{Mint}{HTML}{00B894}

\def\gtrsim{\lower 2pt \hbox{$\, \buildrel {\scriptstyle >}\over
{\scriptstyle \sim}\,$}}
\def\lesssim{\lower 2pt \hbox{$\, \buildrel {\scriptstyle <}\over
{\scriptstyle \sim}\,$}}

\def\approxlt{\lower.2em\hbox{$\buildrel < \over \sim$}}
\def\approxgt{\lower.2em\hbox{$\buildrel > \over \sim$}}

\def\arcmin{$^{\prime}$}
\def\arcsec{$^{\prime\prime}$}

\def\be{\begin{equation}}
\def\ee{\end{equation}}
\def\bea{\begin{eqnarray*}}
\def\eea{\end{eqnarray*}}

\def\arcsec{^{\prime\prime}}
\def\arcmin{^\prime}

\def\jref@jnl#1{{\rm#1}}
\def\aj{\jref@jnl{AJ}}			
\def\araa{\jref@jnl{ARA\&A}}		
\def\apj{\jref@jnl{ApJ}}			
\def\apjl{\jref@jnl{ApJ}}		
\def\apjs{\jref@jnl{ApJS}}		
\def\ao{\jref@jnl{Appl.~Opt.}}		
\def\apss{\jref@jnl{Ap\&SS}}		
\def\aap{\jref@jnl{A\&A}}		
\def\aapr{\jref@jnl{A\&A~Rev.}}		
\def\aaps{\jref@jnl{A\&AS}}		
\def\azh{\jref@jnl{AZh}}			
\def\baas{\jref@jnl{BAAS}}		
\def\jrasc{\jref@jnl{JRASC}}		
\def\memras{\jref@jnl{MmRAS}}		
\def\mnras{\jref@jnl{MNRAS}}		
\def\pra{\jref@jnl{Phys.~Rev.~A}}	
\def\prb{\jref@jnl{Phys.~Rev.~B}}	
\def\prc{\jref@jnl{Phys.~Rev.~C}}	
\def\prd{\jref@jnl{Phys.~Rev.~D}}	
\def\pre{\jref@jnl{Phys.~Rev.~E}}	
\def\prl{\jref@jnl{Phys.~Rev.~Lett.}}	
\def\pasp{\jref@jnl{PASP}}		
\def\pasj{\jref@jnl{PASJ}}		
\def\qjras{\jref@jnl{QJRAS}}		
\def\skytel{\jref@jnl{S\&T}}		
\def\solphys{\jref@jnl{Sol.~Phys.}}	
\def\sovast{\jref@jnl{Soviet~Ast.}}	
\def\ssr{\jref@jnl{Space~Sci.~Rev.}}	
\def\zap{\jref@jnl{ZAp}}			
\def\nat{\jref@jnl{Nature}}		
\def\iaucirc{\jref@jnl{IAU~Circ.}}
\def\aplett{\jref@jnl{Astrophys.~Lett.}}
\def\apspr{\jref@jnl{Astrophys.~Space~Phys.~Res.}}
\def\bain{\jref@jnl{Bull.~Astron.~Inst.~Netherlands}}
\def\fcp{\jref@jnl{Fund.~Cosmic~Phys.}}
\def\gca{\jref@jnl{Geochim.~Cosmochim.~Acta}}
\def\grl{\jref@jnl{Geophys.~Res.~Lett.}}
\def\jcp{\jref@jnl{J.~Chem.~Phys.}}	
\def\jgr{\jref@jnl{J.~Geophys.~Res.}}	
\def\jqsrt{\jref@jnl{J.~Quant.~Spec.~Radiat.~Transf.}}
\def\memsai{\jref@jnl{Mem.~Soc.~Astron.~Italiana}}
\def\nphysa{\jref@jnl{Nucl.~Phys.~A}}
\def\physrep{\jref@jnl{Phys.~Rep.}}
\def\physscr{\jref@jnl{Phys.~Scr}}
\def\planss{\jref@jnl{Planet.~Space~Sci.}}
\def\procspie{\jref@jnl{Proc.~SPIE}}

\def\arcmin{\hbox{$^\prime$}}
\def\arcsec{\hbox{$^{\prime\prime}$}}

\def\ion#1#2{#1$\;${\small\rm\@Roman{#2}}\relax}

\newcount\lecurrentfam
\def\LaTeX{\lecurrentfam=\the\fam \leavevmode L\raise.42ex
\hbox{$\fam\lecurrentfam\scriptstyle\kern-.3em A$}\kern-.15em\TeX}
\def\sizrpt{
(\fontname\the\font): em=\the\fontdimen6\font, ex=\the\fontdimen5\font
\typeout{
(\fontname\the\font): em=\the\fontdimen6\font, ex=\the\fontdimen5\font
}}
\definecolor{dark}{rgb}{ 0.34, 0.09, 0.49}

\usepackage{fancyhdr,url}
\usepackage{transparent}
\usepackage{tablefootnote}
\usepackage{mathpazo}

\begin{document}

\noindent
\begin{raggedright}
{\Huge\bfseries
Chemical Demographics of Star-forming Regions\\[0.5em]
\Large Investigating the universality of complex organic molecules across the protostellar mass spectrum and natal cores}
\end{raggedright}\\

\noindent
Maria N. Drozdovskaya$^{1}$, Timea Csengeri$^{2}$,
Maria Teresa Beltr\'an$^{3}$, Jes J{\o}rgensen$^{4}$,
Alain Baudry$^{2}$, 
Henrik Beuther$^{5}$, 
Laure Bouscasse$^{6}$, 
Laura Colzi$^{7}$, 
Audrey Coutens$^{8}$, 
Marta De Simone$^{9}$, 
Pamela Klaassen$^{10}$, 
Maryvonne Gerin$^{11}$, 
Adam Ginsburg$^{12}$
Daniel Harsono$^{13}$, 
Katharine Johnston$^{14}$, 
Xing Lu$^{15}$, 
S{\'e}bastien Maret$^{16}$, 
Sergio Mart{\'i}n$^{17, 18}$, 
Aina Palau$^{19}$, 
Giovanni Sabatini$^{3}$, 
Miguel Sanz-Novo$^{20}$, 
Maxime Tanious$^{21}$, 
Claudia Toci$^{22}$, 
Yao-Lun Yang$^{23}$, 
Shaoshan Zeng$^{23}$, 
Korash Assani$^{1}$,
Aurore Bacmann$^{16}$, 
Eleonora Bianchi$^{3}$,
Gary Fuller$^{24, 25}$, 
Izaskun Jim{\'e}nez-Serra$^{7}$, 
{\'A}gnes K{\'o}sp{\'a}l$^{26}$, 
Michael K{\"u}ffmeier$^{4}$, 
Álvaro S{\'a}nchez-Monge$^{21, 27}$, 
Fr{\'e}d{\'e}rique Motte$^{16}$, 
Mamta Pandey-Pommier$^{28}$, 
Dries Van De Putte$^{1}$,
V{\'i}ctor Rivilla$^{7}$, 
{\L}ukasz Tychoniec$^{29}$

\begin{tcolorbox}[boxsep=3pt,left=0pt,right=0pt,top=1pt,bottom=1pt,colframe=boxcolor,colback=boxcolor]
\justify
\textbf{Abstract:} Our understanding of complex organic chemistry during the formation of stars and planets is shaped by a handful of cherry-picked sources with bright molecular emission lines. This observational bias impedes our assessment of how universal chemical complexity is during star formation. The lack of systematic wideband spectral surveys for statistically significant samples of protostars and cores is a major roadblock to properly probing how complex organic inventories and abundances depend on the vast range of physical conditions, local cloud and global galactic environments, and evolutionary timescales of low- and high-mass star formation. In turn, this limits our ability to constrain and test physicochemical models, which depend critically on these parameters. The next-generation ALMA2040, with a $10\times$ improvement in line sensitivity, is required to observe complex organics, that we perceive as common today, in a fraction of the observing time that is needed today with ALMA. Such capabilities will propel us into the era of chemical demographics, whereby wideband ($>100$~GHz) spectral surveys at high sensitivity (routinely probing complex organic molecule abundances down to $0.01\%$ relative to methanol) will be executed towards $>1000-10\,000$ prestellar and protostellar sources.

\textbf{With robust chemical inventories towards large source sample sizes, we will be able to probe for dependencies of chemical complexity on mass regime, feedback processes in a cluster, Galactic position, metallicity, and time-variable accretion.} Only with statistically significant samples at high sensitivity will we be able to assess the ubiquity and homogeneity of complex organic chemistry across the protostellar mass spectrum and natal cores. Constraining the diversity and availability of such species is critical for our understanding of the chemical reservoir available for planet formation, which eventually shapes (exo)planetary atmosphere compositions that will be systematically probed with the JWST, Ariel, and the ELT in the coming years. Given the profound significance of complex organics as precursors of prebiotic compounds, systematic homogeneous surveys of these molecules at the onset of star and planet formation is the way to uncover the viability of Solar System-like chemistry in other planetary systems.
\end{tcolorbox}

\begin{comment}
\begin{tcolorbox}[boxsep=3pt,left=0pt,right=0pt,top=1pt,bottom=1pt,colframe=boxcolorapricot,colback=boxcolorapricot]
\justify
\textbf{Key outstanding questions:\\}
 \textbf{1)} Are complex organic molecules ubiquitous and homogeneous across the protostellar mass spectrum and cores? \\
 \textbf{2)} What is the impact of the local cloud and global galactic environment on chemical complexity? \\
 \textbf{3)} Is our Solar System's chemical composition a natural outcome in every low-mass system or a lucky coincidence?
\end{tcolorbox}
\end{comment}

\begin{figure}[b!]
\centering
\includegraphics[trim={0cm 1.0cm 0cm 2.5cm}, clip=true, width=1.0\textwidth]{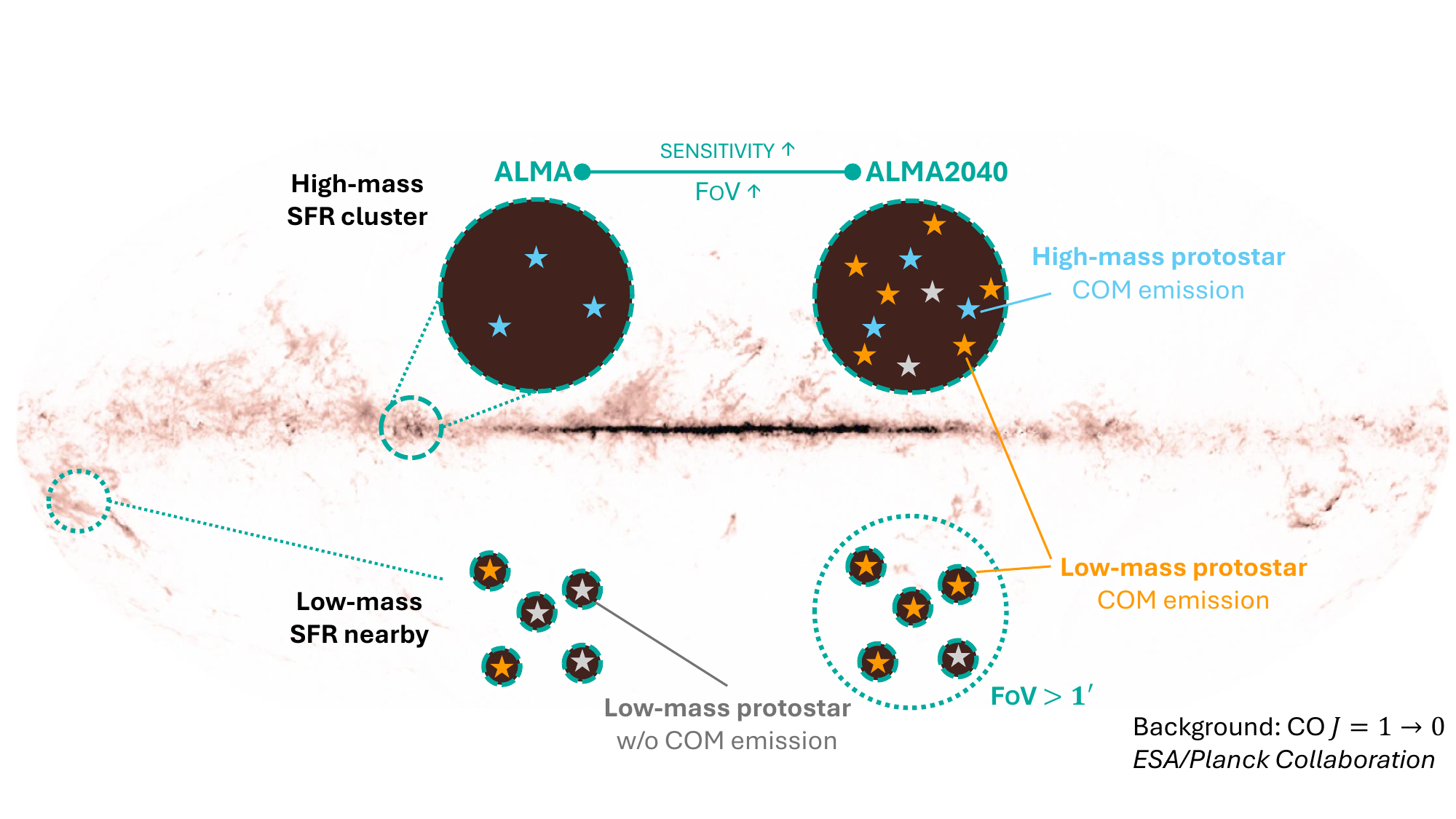}
\caption{\it \textcolor{Gray}{Illustration of ALMA2040 powerful capabilities in comparison to today's ALMA in the context of low- and high-mass star-forming regions (SFRs) and emission from complex organic molecules (COMs).} \vspace{-0.5cm}} \label{fig:depth}
\end{figure}

\newpage{}

\vspace{-0.7cm}
\section{Scientific Context and Open Questions}

\subsection{How universal is chemical complexity in star-forming regions?}
\label{section_COMuniversality}

To date, around 350 different molecules have been discovered in the interstellar medium (Cologne Database for Molecular Spectroscopy, \href{https://cdms.astro.uni-koeln.de}{CDMS}; \citealt{Muller2001, Muller2005}). More than 60\% of these species were detected for the first time in either molecular clouds or star-forming regions \citep{McGuire2022}. However, most of them were discovered towards a handful of high- and low-mass sources (e.g., Sgr B2, W51, G31.41+0.31, Orion KL, TMC-1, IRAS~16293-2422, L483), which have favorable characteristics for detecting complex species, such as high emission line strength, proximity, or narrow line widths. Chemically complex organic species (i.e., C-bearing and composed of at least 6 atoms, \citealt{HerbstvanDishoeck2009}) are found in these sources at abundances of $\sim0.01-1\%$ relative to methanol \citep{Jorgensen2020}, although deep integrations of selected sources can detect larger molecules at even lower abundances. To explore the ubiquity of chemical complexity within the Milky Way, statistically significant sample sizes ($>1000-10\,000$) covering different source populations at an increased  sensitivity are required, which is beyond the realm of current facilities, as they would take years of observing time to complete. \textbf{The preferential study of only the brightest and most prominent sources is insufficient to establish a comprehensive picture of complex organic composition, prevalence, and evolution in the interstellar medium.} Nearby and more distant star-forming regions offer hundreds of low- and high-mass protostars and thousands of prestellar cores: a large sample size suitable for investigating statistical trends among relative molecular abundances and physical parameters spanning several orders of magnitude, such as fundamental protostellar properties (mass, luminosity), environmental conditions, and temporal evolution. Galactic and local natal conditions, and star formation processes set the chemical budget available for subsequent planet formation in protoplanetary disks \citep{ObergBergin2021} and may determine the origin of compounds that contribute to the emergence of life \citep{EhrenfreundCharnley2000}. The atmospheric composition of exoplanets is already being investigated \citep{Tinetti2018,Fu2025}, and by the 2040s, population-level studies of exoplanet atmospheres are fully expected to be available. Overcoming current observational biases is therefore essential for determining the statistical distribution of chemical conditions in star- and planet-forming regions and, ultimately, for linking ALMA (and ALMA2040) disk composition surveys (ALMA2040 white paper by Walsh et al.) to population-level studies of exoplanet atmospheres.

With life on Earth, vinyl cyanide on Titan \citep{Palmer2017}, uracil (one of the four main RNA building blocks) on the asteroid Ryugu \citep{Oba2023}, and glycine (the simplest amino acid) on comet 67P/Churyumov-Gerasimenko \citep{Altwegg2016}, chemical complexity is widespread in the Solar System. Ever more complex and heavier molecules are being regularly discovered in the interstellar medium (e.g., erythrulose, \citealt{Jimenez-Serra2026}), thereby slowly closing the gap between the prebiotic reservoir found in the Solar System and the organic inventory of star-forming regions. However, whether every Sun-like system will host this level of chemical complexity is an entirely open question, because a comprehensive understanding of the prevalent lighter precursor complex organic molecules (e.g., glycolaldehyde) is still lacking. \textbf{Is this type of chemistry ubiquitous in the interstellar medium, or does our selective focus on the richest sources imply a  particular chemistry?} Given that chemical processes are highly sensitive to physical conditions, such as the UV field, gas density, dust and gas temperatures, and the timescales over which these conditions vary \citep{Hasegawa1992, Wakelam2010, Garrod2022}, \textbf{how do the unique environmental conditions at, both, cloud and galactic scales impact the pathways leading to chemical complexity?} It has been suggested that our Solar System formed in a cluster with high-mass stars \citep{PortegiesZwart2009, Adams2010}, but \textbf{what is the role of feedback from high-mass protostars for the complex organic chemistry of neighboring low-mass protostars?} Only large statistical studies can convincingly reveal whether there is diversity or universality when considering complex organic inventories and their relative abundance ratios. ALMA2040, with its extensive chemical surveying capabilities, will transform  our understanding of complex organic chemistry from low- to high-mass star-forming regions with their vastly different prevalent physical conditions and illuminate how unique the life-bearing conditions in our Solar System are.

Obtaining an ``unbiased'' (without prior assumptions on chemical composition) chemical view of large source samples requires instantaneous wideband ($>100$~GHz) spectroscopic observations covering entire atmospheric transmission windows, or multiple bands simultaneously (Table~\ref{tab:sensitivities}). Such wideband capabilities facilitate accurate determination of excitation conditions of small molecules and large complex organics by covering numerous emission lines. ALMA's lower frequencies give access to colder gas in protostellar envelopes and prestellar cores for the most frequently observed complex organic molecules, while higher frequencies give access to warm gas around protostellar sources.  This gives immediate access to the full molecular inventory and the corresponding physical properties of the emitting gas \citep[e.g.,][]{Caux2011, Bouscasse2024}. Such statistical studies would necessarily yield a more complete view of the global chemical composition of star-forming gas in the Milky Way by including significantly fainter and/or colder sources than the archetypal ones. Several heterogeneous (in terms of sensitivity, spatial resolution, emission line coverage) observational programs that study protostellar samples with ALMA currently achieve fractional molecular abundances of complex organic molecules of $>0.01\%$ relative to methanol, at best; however, with significant variation from species to species and source type \citep{Jorgensen2020, Nazari2026}. As an example, the typical methanol column densities reached in survey projects of distant galactic sources are $\sim10^{18}$~cm$^{-2}$, typically yielding a detectable column density threshold of $10^{15}$~cm$^{-2}$ for the most abundant complex organics when the emission is spatially resolved \citep{Chen2023}. For nearby low-mass regions, searches for complex organics with column densities down to a few $10^{14}$~cm$^{-2}$ are possible in sources with methanol column densities of $\sim10^{18}$~cm$^{-2}$ when the emission fills the beam \citep{Jorgensen2018, Coutens2026}. High sensitivity observations with current instruments can stretch these limits to relative abundances down to $0.001-0.0001\%$ with respect to methanol \citep{Mininni2023,Belloche2025}. In order to decipher the origin of chemical complexity in a statistical way, it is mandatory to reach a homogeneous $\sim10^{12}$~cm$^{-2}$ column density sensitivity for a large sample, assuming spatially resolved emission. At these column densities, the most common complex organic molecules would be detected in all types of objects (including those with an intrinsically low luminosity and low gas densities) out to kilo-parsec distances in merely $1$~h of observing time with ALMA2040 (Section~\ref{section_techreq}, Table~\ref{tab:common_COMs}). Large sample sizes of hundreds of sources become available with a typical duration of an ALMA Large Program of $100-200$~h.
 This requires the next-generation facility ALMA2040 to be able to execute such chemical demographics studies by drastically reducing the observing time per source and/or by observing tens of sources simultaneously (Fig.~\ref{fig:depth}). ALMA2040  will be an indispensable cornerstone for a comprehensive interpretation of the hot gas in the innermost regions of protostars, as observed with the ELT, and the cooler enveloping material feeding the terrestrial planet-forming zone accessible at millimeter wavelengths.

\begin{table}[t]
  \centering
  \caption{\textcolor{Gray}{Technical requirements assuming a source at $\delta = -23^\circ\,00'\,00''$, bandwidth of 1~km~s$^{-1}$, dual polarisation, and reference frequencies of 230~GHz and and 90~GHz for Bands~6 and~3, resp.}}
  \label{tab:sensitivities}
  
  \begin{tabular}{llclr}
    \hline
    Science Objective & Facility & Configuration & {Ang. Resolution} & Sensitivity (1$\sigma$) \\
    \hline\hline
    \multicolumn{4}{l}{\textit{ALMA Band 6, 1\,h integration time}} \\
    \addlinespace[2pt]
    Pre-WSU               & ALMA     & C-7, C-8     & 0\rlap{.}{\arcsec}1    &  3.37~K\\
    Low-mass SFR          & ALMA2040 & extended     & 0\rlap{.}{\arcsec}0130 & 21.1~K  \\
    Low-mass SFR          & ALMA2040 & intermediate & 0\rlap{.}{\arcsec}0970 &  0.38~K \\
    Chemical Demographics & ALMA2040 & ---          & 0\rlap{.}{\arcsec}3    &  0.039~K\\
    \midrule
    \multicolumn{4}{l}{\textit{ALMA Band 3, 1\,h integration time}} \\
    \addlinespace[2pt]
    Pre-WSU               & ALMA     & C-8, C-9     & 0\rlap{.}{\arcsec}1    & 24.7~K  \\
    High-mass SFR         & ALMA2040 & extended     & 0\rlap{.}{\arcsec}0301& 23.34~K \\
    High-mass SFR         & ALMA2040 & intermediate & 0\rlap{.}{\arcsec}224  &  0.42~K\\
    Chemical Demographics & ALMA2040 & ---          & 0\rlap{.}{\arcsec}3    & 0.23~K            \\
    \hline
  \end{tabular}
\end{table}

\subsection{Towards unprecedented comprehensive chemical demographics with ALMA2040}
\label{section_chemdemo}

The next-generation ALMA2040 facility could perform unbiased wideband ($>100$~GHz) spectroscopic surveys at high sensitivity for statistically significant source sample sizes by drastically reducing the observing time per source, or by increasing the number of sources observed per pointing with efficient fast mapping techniques (or alternatively, with focal plane arrays, FPAs). Only such observational capabilities are the gateway to studying chemical demographics of star-forming regions. Some of the (known) most abundant O-, N-, S-bearing, and hydrocarbon complex organics are listed in Table~\ref{tab:common_COMs} (e.g., \citealt{Jorgensen2020, Belloche2025}). These molecules are responsible for some of the strongest emission lines in the archetypal sources and will be readily detected with ALMA2040 with its extreme ($10\times$ higher than ALMA) line sensitivity towards much fainter targets. Bandwidths surpassing $100$~GHz will allow such species to be detected in several tens to thousands of rotational transitions, which ensures an accurate determination of excitation conditions and column densities, and is also instrumental for inferring the source structure and serendipitous discovery of new molecular species (ALMA2040 white paper by Rivilla et al.). Interferometric capabilities, in conjunction with the observation of low- and high-energy excited lines across a $>100$~GHz bandwidth, will make it possible to determine precise column densities and distinguish emission from warm ($100-300$~K) and cold ($<100$~K) gas, and to construct spatially resolved column density maps. In the case of particularly bright sources, minor isotopologs of complex organic molecules can be used to constrain the column densities of the main variants and investigate fractionation (ALMA2040 white paper by Colzi et al.). Consequently, ALMA2040 capabilities will allow us to tackle multiple scientific avenues.

\paragraph{\color{Emerald}{\textbullet Are complex organic molecules ubiquitous and homogeneous across the protostellar mass spectrum and cores?}}

The expected $10\times$ increase in line sensitivity at spatial resolutions matching those already reachable with ALMA (e.g., at $0\rlap{.}{\arcsec}1$ or better) enables us to reach about an order of magnitude lower complex organic molecule column densities than achievable today. At the same time, by spatially resolving the differentiated origin of  their emission (such as envelope, disk, outflow), ALMA2040 will tightly constrain the specific chemical processes that govern their formation. For nearby protostars and cores, recovering spatially extended emission is crucial. For distant star-forming regions, a resolution of $0\rlap{.}{\arcsec}3$ can already isolate individual envelopes of hot cores and corinos (resolving disk scales is discussed in the ALMA2040 white paper by K\"{u}ffmeier et al.), thus we can aim for detections at $100\times$ lower column densities than what ALMA currently offers at $0.1\arcsec$ (Section~\ref{section_techreq}). This opens the path to probing complex organics in sources, in which we cannot currently detect complex organics, i.e., many nearby prestellar cores and Class~0/I protostars, and reaching the low- to intermediate mass protostellar population in more distant protoclusters. By being able to probe for complex organics in sources in which we cannot detect such species to date, we will make a major step forward in understanding what the chemical complexity of a ``typical'' star-forming region is and what fraction of star-forming systems naturally produce such molecular inventories.

In the warm inner parts closest to a protostar, complex organic molecules that are initially formed on interstellar dust grain surfaces, thermally desorb at dust temperatures of $100-200$~K. Lack of information on the exact physical structures in these deeply embedded regions that are emitting in complex organic lines resulted in them being collectively called a ``hot core'' and ``hot corino'' in the case of a high-mass and low-mass protostar, respectively. However, emission from complex organics extends beyond these regions, and generally, the phenomena giving rise to complex organic chemistry in pre-protostellar stages remains underexplored on scales of $<0.3\arcsec$. Exploring source structure on smaller spatial scales in conjunction with chemical differentiation will be possible with dedicated follow-up studies for a limited sample. Proposed explanations for the presence of complex organic molecules at dust temperatures well below $100$~K are non-thermal desorption \citep{Bouscasse2024, Busch2024}, gas-phase formation \citep{Balucani2015}, FUV processing \citep{Hernandez-Vera2026}, and various types of shocks \citep{Codella2015, Csengeri2019}. The increased ALMA2040 sensitivity will allow us to probe more accurately what complex organics trace the early warm-up phase  and search for early progenitors of hot cores and hot corinos, which may initially thermally desorb complex organics in only a much smaller region and are entirely missed by current facilities.

Relative ratios between molecular families of complex organic species are valuable tracers of physicochemical processes in star-forming regions. It has also been hypothesized that some of them can be used to determine protostellar ages \citep{Sabatini2021}. Ratios of O- versus N-bearing species would probe potential carbon-grain destruction \citep{vantHoff2020}. Meanwhile, ratios of methanol to ethanol and methyl cyanide to ethyl cyanide would probe the level of progression in chemical complexity. An abundance of unsaturated hydrocarbons with a concurrent lack of O- and N-bearing organics could be interpreted in terms of timescales for core collapse, i.e., statistically verifying the presence of two classes of sources: warm carbon-chain chemistry (WCCC) and hot core- or corino-like \citep{Sakai2009}. Finally, complex S-bearing species could be a prospective window onto the otherwise unobservable semi-refractory component in planet formation \citep{Altwegg2022, Haenni2024}.

\paragraph{\color{Emerald}{\textbullet What is the impact of the local cloud and global galactic environment on chemical complexity?}}

The majority of low- and high-mass protostars, including our Sun, form in clusters \citep{Pfalzner2013, Bergin2024, Beuther2025}. Therefore, it is imperative to investigate the impact of environment and feedback from high-mass stars on the chemical inventory of their accompanying low-mass protostars. Better dynamic range due to the larger number of baselines, as well as the 10$\times$ increase in line sensitivity than what is offered by current facilities is needed to disentangle the bright hot cores and the fainter hot corinos in clustered massive star-forming regions.

Another exciting direction would be the investigation of a complete sample of protostars in a single star-forming region, including those that are located on the edges of a cloud and those in its inner depths, as well as those close to edges of bubbles considered to stem from the triggered mode of star formation. Chemical composition of a molecular cloud is not homogeneous and is driven by its density structure and influenced by external factors, such as proximity to an irradiating source \citep{Aikawa2026, Mazurek2026}. Only a complete cloud-wide chemical census through surveys of the highest column density regions would allow us to investigate if such initial chemical differences propagate to the later protostellar stage.

Statistical samples probing different regions of the Milky Way will probe variations in the chemical composition on Galactic scales. Elemental abundance variations with galactocentric radius is another prime focus of ALMA2040 (ALMA2040 white paper by Colzi et al.); and specific environments at larger galactocentric distances, proximity to the Galactic center, or the Galactic bar may be associated with a significantly different chemical environment (ALMA2040 white paper by Fontani et al.). Hot cores in the outer Galaxy have been identified and similarly show the presence of the most ubiquitous complex organics. The host environment of such cores has a different metallicity and cosmic ray ionization rate, which potentially impact complex organic molecule formation \citep{Shimonishi2021, Wang2026}.

Hot cores in the outer Galaxy have been identified and similarly show the presence of the most ubiquitous complex organics and the potential impact of the different metallicity and cosmic ray ionization rate on the formation of complex organic species has been pointed out \citep{Shimonishi2021, Wang2026}. The effects of environment on the evolution of protoplanetary disks are tackled in the ALMA2040 white paper by Haworth et al. Clearly, statistical studies of large samples across the Galaxy will reveal global galaxy-level trends of the chemistry of complex organic species. Furthermore, a systematic exploration of the brightest hot cores would be achievable beyond a $5$~kpc distance.

\paragraph{\color{Emerald}{\textbullet Is our Solar System’s chemical composition a natural outcome in every low-mass system or a lucky coincidence?}}

From the perspective of its planetary architecture, our Solar System is not exceptionally exotic \citep{Mishra2023, Howe2025}, although observational capabilities down to Earth-sized planets are necessary to draw definitive conclusions about that. What is remarkable is that chemically complex molecules are found all throughout our Solar System on its planets, moons, and minor bodies. However, even next-generation facilities are unlikely to be able to probe for such species in mature exoplanetary systems. Constraining the natal environment of our Solar System and comparing it to other analogous systems is achievable with ALMA2040. It has been demonstrated that the complex organic inventory of comet 67P/Churyumov-Gerasimenko is correlated with that of the low-mass protostar IRAS~16293-2422~B, especially in the CHO-bearing species \citep{Drozdovskaya2019}. This correlation suggests that there is partial inheritance of materials from the earliest stages of star formation to the planetesimals that are eventually formed. Such correlations must be tested for larger samples of protostars and other bodies in our Solar System. Chemical demographics with ALMA2040 will be key in understanding if the chemical complexity of our Solar System is unique.

\begin{table}[t]
    \centering
    \begin{tabular}{lrcc|cc}
        \hline
        Molecule Name & Formula  & $N^{\text{low}}_X$ & $X_{\text{H}_2}^{\text{low}}$ & $N^{\text{high}}_X$ & $X_{\text{H}_2}^{\text{high}}$\\
        & & (cm$^{-2}$) & $N_{\text{H}_2}=10^{23}$ cm$^{-2}$ & (cm$^{-2}$) & $N_{\text{H}_2}=10^{25}$ cm$^{-2}$\\
        \hline\hline
        \textbf{O-bearing} & & \\
        \hline
        Methanol & CH$_{3}$OH &  $7\times10^{13}$ & $7\times10^{-10}$ & $2\times10^{14}$ & $2\times10^{-11}$   \\
        Methyl formate & CH$_{3}$OCHO  &  $5\times10^{13}$ & $5\times10^{-10}$ & $1\times10^{14}$ & $1\times10^{-11}$\\
        Acetaldehyde & CH$_{3}$CHO  & $3\times10^{13}$ & $3\times10^{-10}$ & $7\times10^{13}$ & $7\times10^{-12}$ \\
        Dimethyl ether & CH$_{3}$OCH$_{3}$ & $6\times10^{13}$ & $6\times10^{-10}$ & $2\times10^{14}$ & $2\times10^{-11}$\\
        Ethanol & C$_{2}$H$_{5}$OH & $1\times10^{14}$ & $1\times10^{-9}$ & $3\times10^{14}$ & $3\times10^{-11}$\\
        \hline
        \textbf{N-bearing} & & \\
        \hline
        Formamide & NH$_{2}$CHO &   $4\times10^{12}$ & $4\times10^{-11}$ & $2\times10^{13}$ & $2\times10^{-12}$\\
        Methyl isocyanate & CH$_{3}$NCO &  $5\times10^{13}$ & $5\times10^{-10}$ & $1\times10^{14}$ & $1\times10^{-11}$ \\
        Methyl cyanide & CH$_{3}$CN &  $1\times10^{12}$ & $1\times10^{-11}$ & $4\times10^{12}$ & $4\times10^{-13}$\\
        Ethyl cyanide & C$_{2}$H$_{5}$CN &  $9\times10^{12}$ & $9\times10^{-11}$ & $3\times10^{13}$ & $3\times10^{-12}$ \\
        Vinyl cyanide & C$_{2}$H$_{3}$CN &  $9\times10^{12}$ & $9\times10^{-11}$ & $2\times10^{13}$ & $2\times10^{-12}$  \\
        \hline
        \multicolumn{2}{l}{\textbf{Hydrocarbons \& S-bearing}} &   \\
        \hline
        Propyne & CH$_{3}$CCH &  $4\times10^{13}$ & $4\times10^{-10}$ & $1\times10^{14}$ & $1\times10^{-11}$\\
        Methyl mercaptan & CH$_{3}$SH & $6\times10^{13}$ &  $6\times10^{-10}$ & $1\times10^{14}$ & $1\times10^{-11}$ \\
        \hline
    \end{tabular}
    \caption{\textcolor{Gray}{Molecular column density detection thresholds with $1$~h of integration time in ALMA's Band~6 for a 0\rlap{.}{\arcsec}3 synthesized beam size using Local Thermodynamic Equilibrium (LTE) models assuming spatially and spectrally resolved emission for a $150$~K gas kinetic temperature and a line width of $1$~km~s$^{-1}$ and $3$~km~s$^{-1}$ for the low- and high-mass case, resp. The detection limit corresponds to at least two transitions reaching a peak brightness temperature above $3\sigma$, where 1$\sigma$ is 0.039~K and 0.23~K, resp (Table~\ref{tab:sensitivities}). Only transitions in the vibrational ground state are considered.}}
    \label{tab:common_COMs}
\end{table}

\vspace{-0.5cm}
\section{ALMA2040: Technical Requirements}
\label{section_techreq}

The current wideband capabilities of (sub)millimeter interferometers are driven by the SMA ($32$~GHz single polarisation) and NOEMA ($32$~GHz dual band, dual polarisation), while ALMA WSU ($16$~GHz dual polarisation)\footnote{The new generation correlator (ATAC) is capable of processing $32$~GHz in dual polarization; however, its implementation is contingent on available funding.} is already on the horizon to improve the array's current spectroscopic capabilities. Capturing an instantaneous bandwidth of 20~GHz and above is already technically feasible and is expected to reach
$>100$~GHz in the future, thereby covering a large fraction of the atmospheric transmission windows. 
ALMA2040 will thus significantly increase the instantaneously observable bandwidth together with a major increase in sensitivity thanks to the large collecting area (offered by its $200-300$ antennas).

To enable chemical demography, different source populations need to be sampled. As frequency coverage and angular resolution can be jointly optimized according to source distance, morphology, and the physical scales of interest, we discuss below two suitable observing regimes for these populations. Should a $>$100~GHz instantaneous bandwidth be available, these could be exploited as distinct sub-regimes. Our science case requires an angular resolution of $\sim0.3\arcsec$, which corresponds to $150$~au at $0.5$~kpc and $900$~au at $3$~kpc. For nearby sources, this allows the inner star-disk region to be disentangled from the envelope, thus facilitating spatially resolved studies of individual complex organics. For sources at distances of $<3$~kpc, we would be able to resolve the majority of individual cluster members in the high-mass regime (as shown in \citealt{Bonfand2024}), paving the way to the exploration of fragmentation and disk properties on scales $<0.3\arcsec$ (ALMA2040 white papers by Sanchez-Monge et al. and K\"{u}ffmeier et al.). An angular resolution of $\lesssim0.3\arcsec$ is also instrumental in reducing line blending, as it isolates individual emitters spatially and reduces the number of different velocity components observed in one beam.

At typical gas temperatures of hot corinos and hot cores ($100-300$~K), line emission of several of the complex organic molecules discussed here peaks in the $200-300$~GHz range, which makes ALMA's Band~6 one of the most suitable frequency ranges for such studies. However, as line density decreases at lower frequencies, ALMA's Band~3 can be advantageous for particularly bright, line-rich sources in high-mass star-forming regions due to lower line confusion. We provide illustrative numbers for one setup covering the entire Band~6 and another aimed at lower frequencies covering Band~3 in Table~\ref{tab:sensitivities}\footnote{Assuming $>100$~GHz instantaneous bandwidth, this corresponds to ALMA WSU's Bands~2 to 4.}. However, ALMA's Bands 1 through 8 (including those to be upgraded within the scope of the WSU) are all equally important, because a wide range of frequencies facilitates observations of lines across a broad range of upper level energies, which is key for accurate excitation temperature and column density constraints. Moreover, if a bandwidth much larger than 100 GHz will be offered, it will be possible to concurrently observe several lines of smaller precursor species (such as HCN, H$_{2}$S, OCS, PN) that have few, sparsely spaced transitions. In turn, this will allow their excitation temperature and column densities to be robustly determined, which currently requires multi-band observations in many tunings and remains challenging for today's ALMA and even the planned WSU upgrades.

ALMA2040 with its larger collecting area ($200-300$ antennas) represents a major increase in sensitivity. A 1~h long on-source integration yields a surface brightness sensitivity of $0.38$~K at $1$~km~s$^{-1}$ at Band~6 in a $0.1\arcsec$ beam, which is about $10\times$ better line sensitivity compared to today's ALMA, and captures at least one order of magnitude lower abundances for complex organic species. However, relaxing this angular resolution by a factor of $\sim3$ would still allow us to disentangle dominant sources and yet, yield an overall gain of nearly two orders of magnitude in sensitivity and probe two orders of magnitude lower complex organic abundances. Overall, this would allows us to reach $3\sigma$ detections for molecular column densities ($N$) on the order of $10^{12}-10^{13}$~cm$^{-2}$ for several complex organic species towards low-mass star-forming regions assuming spatially and spectrally resolved emission in 1~h of integration time in Band~6. Assuming $N_{\rm{H_2}}\sim10^{23}$~cm$^{-2}$ that is typical for low-mass protostellar envelopes, relative molecular abundances ($X$) of about $10^{-10}-10^{-9}$ can be reached. Due to their intrinsically larger line widths (due to turbulence and blended gas motions), this sensitivity limit translates to somewhat higher detectable complex organic molecule column densities, typically on the order of $10^{13}-10^{14}$~cm$^{-2}$ for their high-mass equivalents. However, assuming $N_{\rm{H_2}}\sim10^{25}$~cm$^{-2}$ that is typical for high-mass protostellar envelopes, the attainable relative molecular abundances reach a lower regime of about $10^{-12}-10^{-11}$ (Table~\ref{tab:common_COMs}). At such high column densities, the dust may become optically thick and obscure the line emission, or drive it into absorption. Inclusion of low frequencies in the wideband spectral surveys becomes key here, as such dust opacity effects increase with frequency and so can then be directly observationally probed.

Our current statistics yields 4-10 chemically rich sources per pointing for protoclusters at distance of up to $5$~kpc \citep{Bonfand2024}; however, considering the sensitivity to lower abundances means that the statistics of sources with complex organic molecule detections could be even higher. Multiple sources per pointing are also expected towards nearby cluster-forming regions; however, more efficient mapping capabilities would be required to cover a larger field of view (FoV). Fast fringe tracking would greatly improve our current mapping capability of large mosaicked areas. Alternatively, this could be achieved with FPAs. Simultaneously observed large FoVs of a few arcmin$^{2}$ would allow studies of all subregions of entire protoclusters.

A higher sensitivity allows us to simultaneously capture chemically rich and poor, as well as hot and cold sources within each field, providing a good compromise between the expected number of sources and the required integration time. With 1~h of integration time per field, each field containing 4-10 sources with complex organic molecule emission (including those in which today's instruments do not detect such emission), a sample size of 100s to 1000 of sources becomes observable within $<300$~h of integration time at significantly higher sensitivity.

As already planned for the ALMA WSU, a spectral resolution reaching $0.1$~km~s$^{-1}$ would be required, driven by the need to spectrally resolve the narrower line widths typical of prestellar cores ($\sim0.3$~km~s$^{-1}$) and quiescent parts of low-mass protostellar envelopes ($\sim1$~km~s$^{-1}$). For precise estimates of molecular column densities, molecular lines need to be spectrally resolved, which requires typically at least three resolution elements per line. For a systematic study of, both, low- and high-mass sources with line widths in the $\sim1-5$~km~s$^{-1}$ range, spectral smoothing to the lowest common line width could yield improved sensitivities.

\vspace{-0.5cm}
\section{Synergies with Current and Future Facilities}
\label{section_otherfacil}

In the 2040s, when ALMA2040 would become operational, several high angular resolution observatories will be in operation across the electromagnetic spectrum.

\vspace{-0.5cm}
\paragraph{\color{Emerald}{\textbullet~~Extremely Large Telescope (ELT)}}

The ELT will probe the innermost terrestrial regions of planet-forming disks and measure the atmospheric compositions of gaseous and rocky exoplanets in near- and mid-infrared at $12-53$~mas angular resolution. Chemical demographics with ALMA2040 would fill in the missing information on the chemical composition of the material initially available for forming protoplanetary disks and planets.

\vspace{-0.5cm}
\paragraph{\color{Emerald}{\textbullet~~{Square Kilometer Array (SKA)-Mid}}}

SKA's FoV of $\sim7\arcmin$ at $15$~GHz will capture multiple protostellar sources with every pointing, which means that it is critical for ALMA2040 to follow up many sources in short observing times. SKA's cm wavelengths will be complementary to ALMA2040, especially for sources in which large dust optical depth suppresses complex organic molecules line emission at mm wavelengths \citep{DeSimone2020, Bianchi2026}. The overlap of SKA's Band 6 and ALMA's Band 1 would be extremely powerful in searches for the heaviest complex organics in cold sources.

\vspace{-0.5cm}
\paragraph{\color{Emerald}{\textbullet~~Next Generation Very Large Array (ngVLA)}}

The ngVLA's frequency coverage is optimal for detecting prebiotic compounds in cold sources \citep{McGuire2018_ngVLAprebio}. Chemical demographics with ALMA2040 would be pivotal for identifying sources with the highest precursor abundances and most suitable spectral characteristics (e.g., line strength and width) for deep follow-up ngVLA observations. If ngVLA reaches the envisioned $116$~GHz frequencies, then there is an opportunity for full sky coverage in the $30-116$~GHz regime with ngVLA in the North and ALMA2040 in the South. ngVLA opens complex organic studies in dust optically thick sources at mm wavelengths.

\vspace{-0.5cm}
\paragraph{\color{Emerald}{\textbullet~~James Webb Space Telescope (JWST)}}

Although concurrent operation of JWST and ALMA2040 may be limited, two decades of JWST operations will yield statistically significant insights on the chemical composition of interstellar ices and hot gas. ALMA2040 will then recover the missing information on cold gas \citep{Chen2024, Gieser2026}. JWST extreme sensitivity and Fov of several arcminutes facilitate identification of numerous new protostars in highly extincted environments, yielding new targets for chemical demographic studies. Chemical characterization of exoplanetary atmospheres with JWST requires interpretation in the context of their natal regions.

\vspace{-0.5cm}
\paragraph{\color{Emerald}{\textbullet~~ESA Voyage 2050}}

With the selection of Saturn's moon Enceladus as the target of the next L4 mission, ESA made a major commitment to understanding the origin of life. Complex organic molecules in star- and planet-forming regions could be the initial feedstock of prebiotic compounds that arrive on a planetary surface. Therefore, understanding chemical complexity of a broad range of infant star- and planet-forming regions must be a top priority for ALMA2040.

The ambitious observing program of this white paper will pioneer the era of chemical demographics of star-forming regions. It will provide an impactful, long-term legacy value for the community, serving as a homogeneous referential database for, both, data mining by observers using stacking and various AI methods for deep searches of specific species, as well as for astrochemical modelers in their pursuit of constraining the vast physicochemical parameter space of interstellar complex organic chemistry. Complex organic demographics of the earliest stages of star and planet formation will serve as a benchmark for understanding the chemistry of ensuing planet formation. A coherent survey of chemical complexity across our Milky Way will facilitate the interpretation of extragalactic complex organic detections.

\noindent\textit{\textcolor{Gray}{This work was developed as part of the broader ALMA2040 community effort. We acknowledge the Lorentz Center and NOVA for supporting the community workshops and coordination activities that helped initiate and shape the ALMA2040 White Paper process. MND's work is funded by the European Union. Views and opinions expressed are however those of the author(s) only and do not necessarily reflect those of the European Union or the European Research Council Executive Agency. Neither the European Union nor the granting authority can be held responsible for them. MND's work is supported by ERC grant PSII (DOI: 10.3030/101230593). TCs acknowledges support from the Thematic Actions ``Physique et Chimie du Milieu Interstellaire'' (PCMI) and ``Action Th{\'e}matique Physique Stellaire'' (PS) of INSU Programme National ``Astro'', with contributions from CNRS Physique \& CNRS Chimie, CEA, and CNES.}}

\bibliographystyle{apj_mod}
{\bibliography{bibliography}}

\newpage
\section*{Affiliations}

\vspace{1em}
{
\footnotesize{
\noindent~$^{1}$~Department of Chemistry, Biochemistry and Pharmaceutical Sciences (DCBP), Universit\"{a}t Bern, Freiestrasse 3, 3012 Bern, Switzerland \\
$^{2}$~Laboratoire d’Astrophysique de Bordeaux, Universit\'{e} de Bordeaux - CNRS, Bat. B18N, Allee Geoffroy Saint-Hilaire, 33600 Pessac Cedex, France \\
$^{3}$~INAF-Osservatorio Astrofisico di Arcetri, Italy\\
$^{4}$~Niels Bohr Institute, University of Copenhagen, Denmark\\
$^{5}$~Max Planck Institute for Astronomy (MPIA), Germany\\
$^{6}$~Institut de Radioastronomie Millim\'{e}trique (IRAM), Saint-Martin-d'H\`{e}res, France\\
$^{7}$~Centro de Astrobiolog{\'i}a (CAB), CSIC-INTA, Spain\\
$^{8}$~Univ. Toulouse, CNES, CNRS, IRAP, Toulouse, France\\
$^{9}$~European Southern Observatory (ESO), Germany\\
$^{10}$~UK Astronomy Technology Centre, Royal Observatory Edinburgh, UK\\
$^{11}$~LUX, Observatoire de Paris, PSL Research University, CNRS, Sorbonne Universit{\'e}, France\\
$^{12}$~University of Florida, USA\\
$^{13}$~Institute of Astronomy, Department of Physics, National Tsing Hua University, Taiwan\\
$^{14}$~University of Lincoln, UK\\
$^{15}$~Shanghai Astronomical Observatory, China\\
$^{16}$~Univ. Grenoble Alpes, CNRS, IPAG, France\\
$^{17}$~European Southern Observatory (ESO), Chile\\
$^{18}$~Joint ALMA Observatory, Chile\\
$^{19}$~Instituto de Radioastronomia y Astrofisica, Universidad Nacional Autonoma de Mexico, Mexico\\
$^{20}$~Max-Planck-Institut f{\"u}r extraterrestrische Physik (MPE), Germany\\
$^{21}$~Institut de Ci\`{e}ncies de l'Espai (ICE-CSIC), Spain\\
$^{22}$~Departamento de Fisica Aplicada III, Universidad de Sevilla, ETSI, Spain\\
$^{23}$~RIKEN, Japan\\
$^{24}$~Jodrell Bank Centre for Astrophysics, UK\\
$^{25}$~I. Physikalisches Institut, Universit{\"a}t zu K{\"o}ln, Germany\\
$^{26}$~Konkoly Observatory, HUN-REN Research Centre for Astronomy and Earth Sciences, MTA Centre of Excellence, Hungary\\
$^{27}$~Institut d’Estudis Espacials de Catalunya (IEEC), Spain\\
$^{28}$~University Catholic of Lyon, University of Lyon, France\\
$^{29}$~Leiden Observatory, Leiden University, The Netherlands\\
}
}

\end{document}